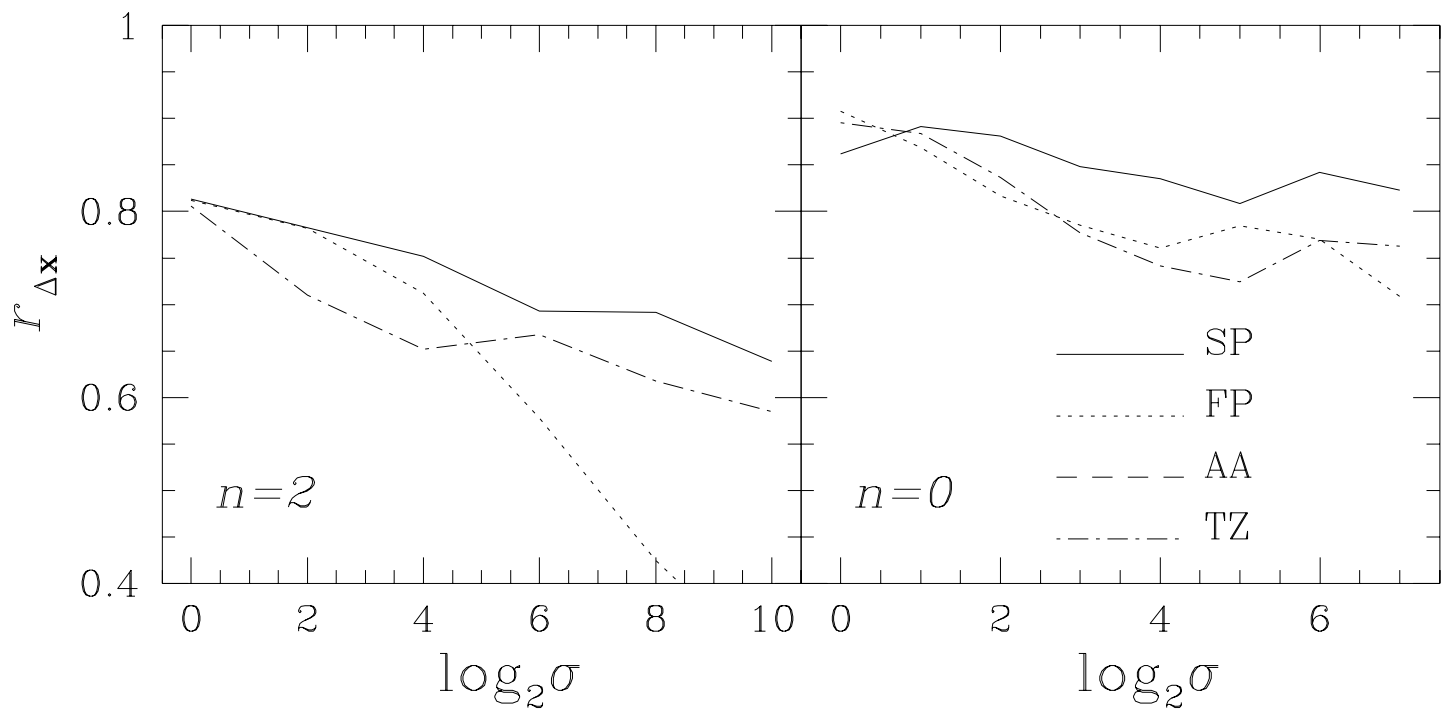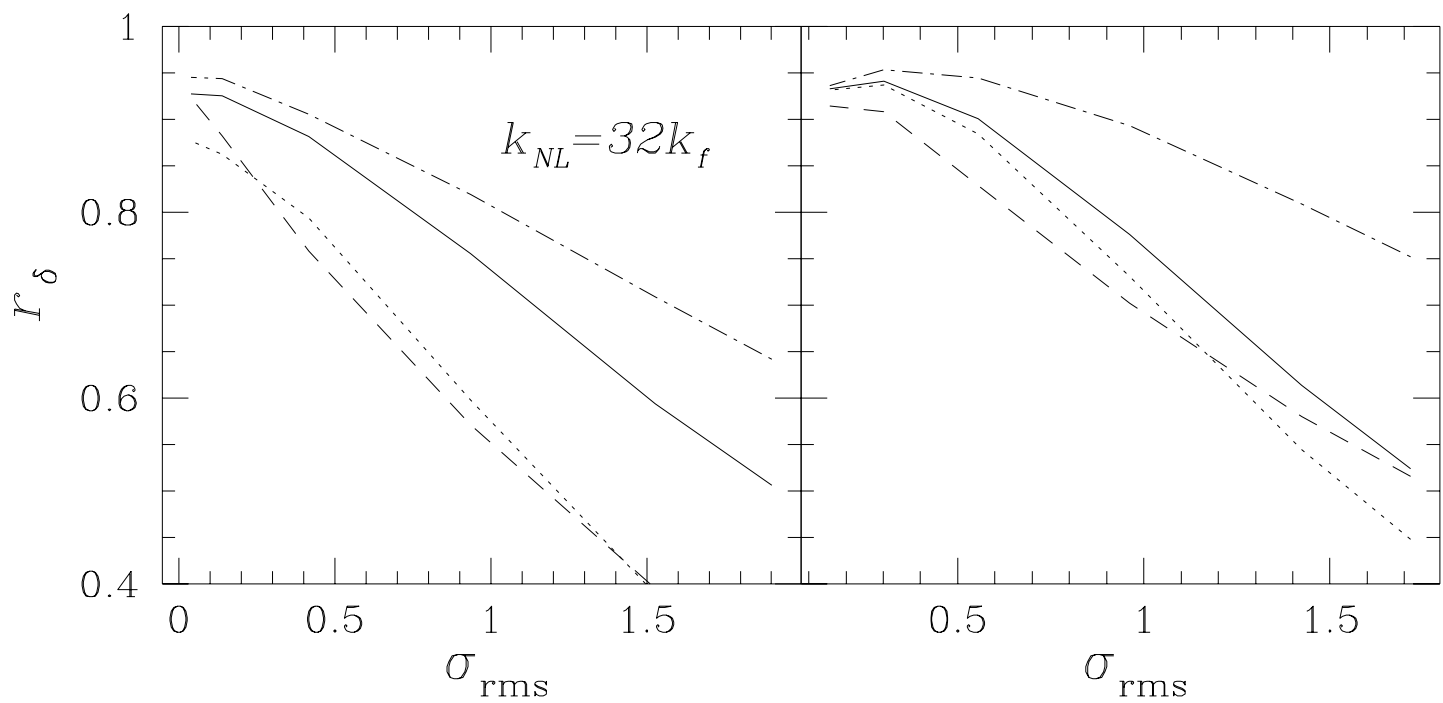

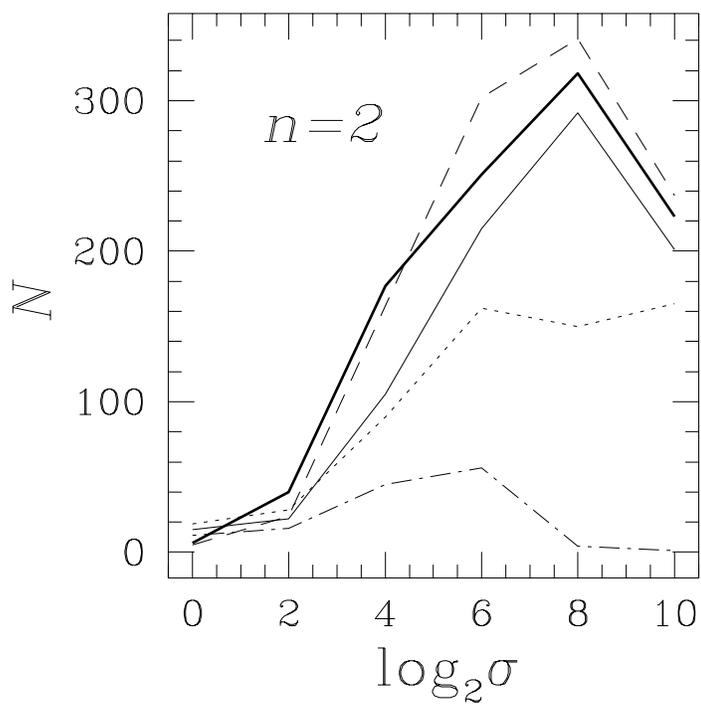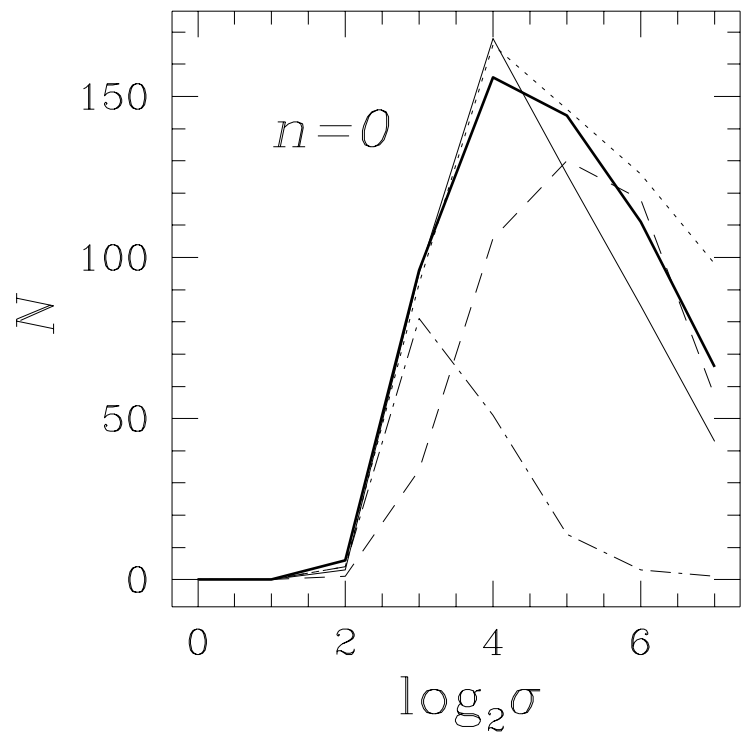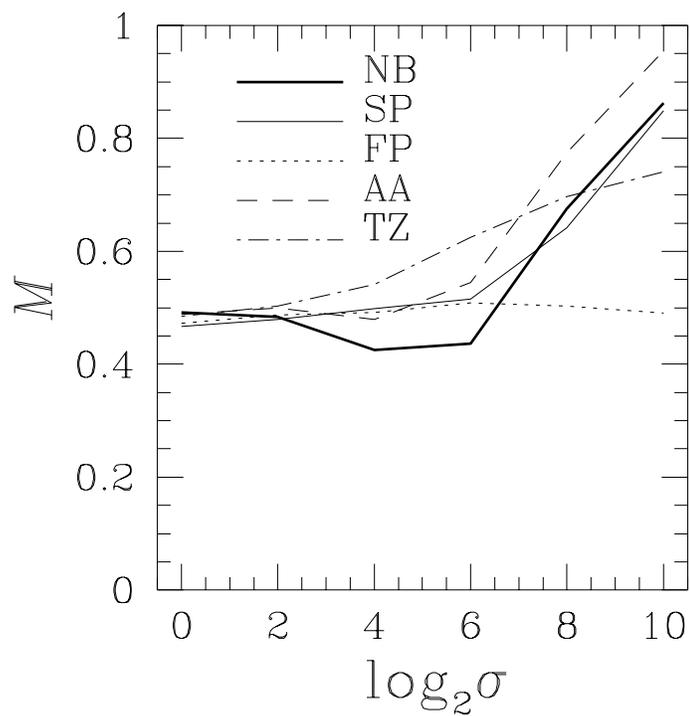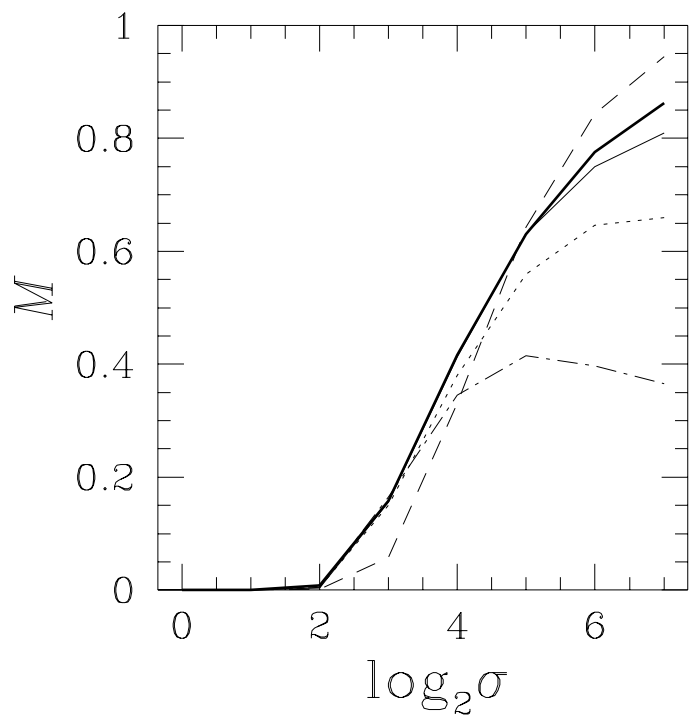

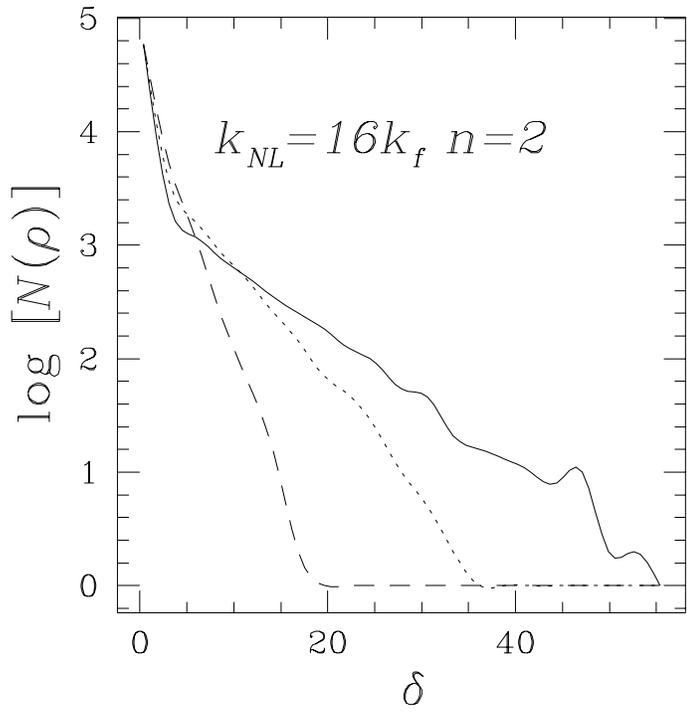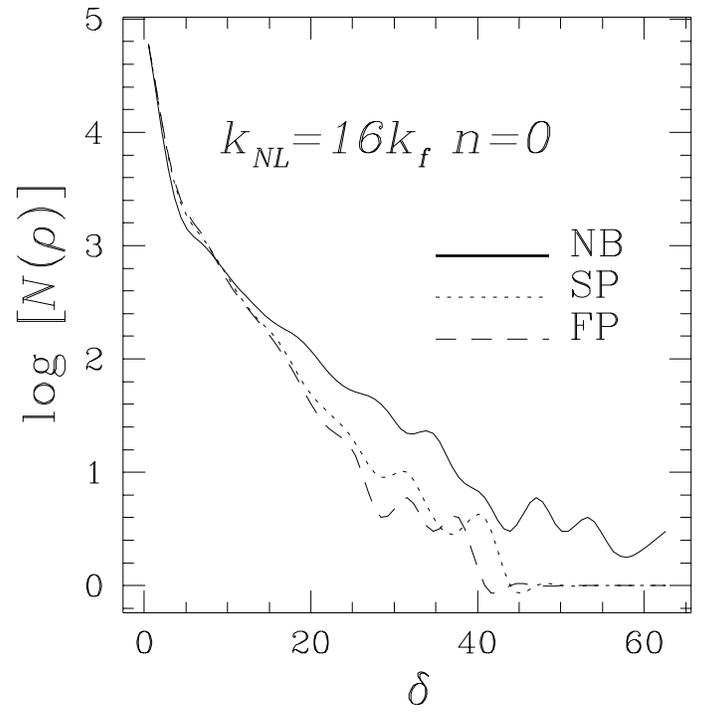

# Evolution of the Potential in Cosmological Gravitational Clustering

Adrian L. Melott

Department of Physics and Astronomy, University of Kansas, Lawrence, KS 66045

B.S. Sathyaprakash and Varun Sahni

Inter-University Centre for Astronomy and Astrophysics, Post Bag 4, Ganeshkind, Pune 411 007 India

## ABSTRACT

The gravitational potential is a constant to linear order in cosmological gravitational clustering. In this Letter we present results of testing the conjecture, proposed by Pauls and Melott (1995), that the effect of nonlinear evolution on the potential can be better described by smoothing it on the scale of nonlinearity. We show two-dimensional simulations consisting of an N-body code in which particles are accelerated not by their mutual attraction, but by the gradient of the initial potential smoothed on the current scale of nonlinearity. This approximation produces results considerably improved over using a constant potential to move particles, and it is generally better than any scheme we have tested, lending further support to the smoothing description of the evolved potential.

## 1. INTRODUCTION

In recent years, there has been an increasing interest in gravitational instability, partly motivated by a consensus that it is the primary driving process for the formation of structure in the Universe. There has especially been a great deal of progress in our understanding of the mildly nonlinear regime of gravitational clustering (for a review see Sahni and Coles 1995). Systematic testing and comparison has led to the conclusion that Lagrangian-based schemes, closely related to the Zel'dovich (1970) approximation (ZA) are much superior to others (e.g. Melott 1994; Munshi, Sahni and Starobinsky 1994; Sathyaprakash et al. 1995, hereafter SSMPM). More recent work has focused on understanding why schemes based on ZA work so well outside their original realm of applicability (which was the class of models with truncated initial power spectra called 'pancake models') and also on finding better approximations. The understanding of quasilinear and nonlinear clustering is important to explain the so-called filamentary structure observed in the Universe, to our understanding of gravitational instability theory in general, and to the development of numerical or analytic schemes which can eliminate the necessity for N-body simulations for certain needs and/or permit the rapid creation of large ensembles of realizations of a given scenario.

In this Letter we present new results on an approximation based on a new description of the evolution of the gravitational potential. The Newtonian limit of General Relativity is used in cosmological regimes where curvature and speeds do not invalidate it. The potential is much more dominated by long waves than other quantities such as density or velocity fields, so it evolves more slowly. It also generalizes more easily to GR. Based on this Brainerd, Scherrer and Villumsen (1993) and Bagla and Padmanabhan (1994) proposed an approximation called 'Frozen Potential' (FP). FP consists of performing an N-body simulation but *without* recalculating the potential which is constant to linear order in a critical density universe. These groups tested the approximation on some spectra and concluded it worked well. However, they did not look at a wide variety of spectra nor did they compare it with many currently preferred approximations. SSMPM conducted such a comparison and found that while it did have a range of validity, FP was less accurate than some other schemes which were easier to apply. One such example is the Truncated Zel'dovich Approximation (TZ) which consists of smoothing the initial conditions before applying the Zel'dovich approximation (e.g. Coles, Melott and Shandarin, 1993; Melott, Pellman and Shandarin, 1994).

Pauls and Melott (1995) have examined the behavior of TZ in more detail, showing how it can (for example) predict the orientation of superclusters from initial conditions. They also showed evidence for a more complete description of the evolution of the potential: the potential of the initial conditions, smoothed on the scale of nonlinearity, more closely resembles the evolved N-body potential than does the unsmoothed initial potential. Potentials were smoothed by convolution with a Gaussian $e^{-k^2/2k_G^2}$ where $k_G \sim k_{\rm NL}$ defined by

$$a^2 \int_0^{k_{\rm NL}} d^D k P(k) \equiv 1. \qquad (1)$$

Here $P(k)$ is the initial ($a = 1$) power spectrum of mass density perturbations and $D$ is the dimensionality of space. Even with power-law spectra $P(k) \propto k^n$, with $n = -1$, improvement was found; the improvement was greater for more positive $n$.

## 2. THE STEPWISE SMOOTHED POTENTIAL APPROXIMATION

In this letter, we propose and show encouraging first results on a new approximation we call Stepwise Smoothed Potential (SP). In spirit, this is similar to FP, but at each time step the initial potential is successively smoothed on a larger scale, corresponding to the growth of nonlinearity according to Equation (1). For the purpose of smoothing we employ a Gaussian window $e^{-k^2/2k_G^2}$, where $k_G^{-1}$ is the smoothing scale. No information about the movement of particles is fed back into the gravitational field. This constitutes a more extended test of the smooth potential ansatz proposed by Pauls and Melott (1995).

We used for comparison the high-resolution two dimension PM simulations described in Beacom et al. (1991), which were the basis of extensive tests on many approximations by SSMPM. That study concluded that TZ and another approximation called the adhesion approximation (AA) (Shandarin and Zel'dovich 1989; Kofman et al. 1992; Sahni et al. 1994) performed best in most respects. We have repeated all the tests conducted in SSMPM but as there is insufficient space here for all our results, we shall show only the most direct tests of agreement.

In this study we examine two-dimensional power-law models $n = 2$ and $n = 0$, analogous to $n = +1$ and $n = -1$ in three dimensions. We also examined $n = -2$ (equivalent to $n = -3$ in three dimensions) but found no particular improvement in smoothing the potential. This result is expected since the gravitational potential evolves much more slowly in models such as $n = -2$ which have substantial large-scale power. In this study we varied $k_G$ in units of 0.5 $k_{\rm NL}$ and found the best agreement (as defined by crosscorrelation of resulting density fields) for $k_G \sim 1.5 k_{\rm NL}$. (Of course, $k_{\rm NL}$ changes with time, so we had to recalculate the potential by smoothing at each timestep. Thus, SP is *not* a particularly fast method, and is not as good as TZ is for rapidly generating realizations.)

Our first, and admittedly qualitative comparison is in Fig. 1, where we show dot plots of N-body simulations (middle panels) and the results of SP (top panels) together with those of FP (bottom panels), for the $n = 0$ and $n = 2$, models as labeled. It can be seen that the agreement between N-body and SP is quite good, even for the particularly challenging $n = 2$ case. In the $n = 0$ case too SP shows improvement over FP. Comparison with plots in SSMPM should convince the reader that the agreement is as good as for any of the plots shown there.

Quantitative tests of resemblance begin in Fig. 2. The top panels check particle displacement. If $\mathbf{\Delta X}_Q^i$ ($\mathbf{\Delta X}_N^i$) is the vector displacement of particle $i$ from the initial position by approximation Q (respectively, N-body), then we define the vector correlation coefficient of particle positions by

$$r_{\mathbf{\Delta X}} = \frac{\mathbf{\Delta X}_Q^i \cdot \mathbf{\Delta X}_N^i}{\left[\left(\mathbf{\Delta X}_Q^i \cdot \mathbf{\Delta X}_Q^i\right)\left(\mathbf{\Delta X}_N^j \cdot \mathbf{\Delta X}_N^j\right)\right]^{1/2}}. \qquad (2)$$

The quantity $r_{\mathbf{\Delta X}}$, which measures the agreement in the displacements, is shown plotted versus $\sigma$, the linear theory RMS density contrast at the Nyquist frequency limit. We show for comparison results from SSMPM for the closely related FP, as well as the previously successful TZ and, where applicable, AA. Clearly, by this measure, SP is unambiguously the best. (In this case AA is not included because our solution for AA does not push particles.)

The bottom panels show the mass density crosscorrelation coefficient

$$r_\delta = \frac{\delta_Q^i \delta_N^i}{\left\langle (\delta_Q^i \delta_N^i)^2 \right\rangle^{1/2}} \qquad (3)$$

where $\delta^i$ is the density contrast in cell $i$. This approaches unity for perfect agreement. All density fields (at stage $k_{NL} = 32 k_f$) were smoothed using the same variable Gaussian filter. In this case SP is not the best. TZ remains the best due to its precision placement of larger structures. Density fields computed at very high resolution yield very low correlation coefficients even when the errors in the positions of mass concentrations are tiny. Behavior of $r_\delta$ (cf. equation (3)) when different smoothings are employed can tell us how well a given approximation scheme forms mass concentrations on different scales.

As measured by this test, SP performs better than any approximation except TZ. However, the comparison with AA is somewhat ambiguous because our AA solution method does not specify any mass distribution inside clumps; it had to be generated by an ad hoc smoothing procedure described in SSMPM. SP again shows improvement over FP, reinforcing the improved description of the evolving potential.

Next, we show data on clumps of matter. We define a clump as a connected region of overdensity above a certain threshold density $\rho_c$ (see figure caption). Top panels in Fig. 3 show the evolution of the number of clumps $N$ as a function of $\sigma$. The mass of clumps $M$ (bottom panels) is simply the total mass of these objects as a fraction of the mass in the simulations. Fig. 3 shows that SP is a major improvement over all others including AA. Note especially that in the $n = 2$ case, where we see hierarchical clustering and the consequent fall-off in the number of clumps at later epochs, FP does not predict merger of clumps, while SP follows the N-body curve quite accurately. We have tested the agreement between SP and N-body by choosing different density thresholds $\rho_c$ and found the agreement to be quite insensitive to our subjective definition of clumps.

Finally, in Fig. 4 we show the number of cells in our simulations with density contrast in the range $\delta$ and $\delta + d\delta$, which is an estimate of the probability distribution of the density field. Here again we see a substantial improvement in SP over FP in both models. It is clear that by successively smoothing the potential it is possible to form high density clumps just as in N-body simulations. Note, however, that there is a clear indication that just smoothing the potential is insufficient to form structures of very high density contrast.

## 3. CONCLUSIONS

To summarize, on the whole the stepwise smoothed potential is the most accurate of any of the many approximations we have tested. However, we do not suggest its direct use in generating approximate model Universes. It shares the fatal flaw with FP that it is nearly as time-consuming as a full PM simulation (although it would be easier to parallelize). Nor does it have analytic solutions as do AA and TZ. The real importance of its remarkable accuracy, and its substantial improvement over FP, lies in the evidence for the description of the potential as evolving primarily by smoothing.

This result can be understood by a simple physical picture. Consider a condensation sitting at the center of a region in comoving coordinates, which we approximate as spherical, from which its mass originated. The potential on that spherical surface would be the same as if all structure inside the sphere had been erased in the initial conditions.

This is of course not a complete description of the evolution of the potential. There should be deep cusps at dense objects (close examination reveals somewhat diffuse objects in SP) but this does not much affect the motion of these clumps. SP clearly works well compared with many other approximations, and is a major improvement over FP, validating the description by Pauls and Melott (1995). It may find a practical application in providing enhanced boundary conditions for simulations.

This behavior of the potential also helps to account for the unreasonable effectiveness of TZ, which was originally applied to initial conditions with damped (high frequency cutoff) power spectra, as well as SP. Clumps have at any moment largely virialized power on frequencies $k > k_{\rm NL}$, and are moving coherently in a smooth potential. Merging processes intrinsic to hierarchical clustering are driven by power at smaller $k$, and naturally resemble motions under SP.

**Acknowledgements:** We acknowledge useful conversations with Peter Coles, David Weinberg, and participants in the 1994 Aspen Cosmology workshop, as well as financial support form NASA grant NAGW-3832 and computing at the National Center for Supercomputing Applications, Urbana, Illinois.

---

This preprint was prepared with the AAS LATEX macros v3.0.

Fig. 1.— Plots showing particle positions at an epoch when $k_{\rm NL} = 8k_f$ corresponding to SP (top panels) NB (middle panels) and FP (bottom panels). The left-hand (right-hand) panels correspond to the $n = 2$ ($n = 0$) model.

Fig. 2.— Evolution of the vector correlation coefficient of particle positions (top panels). Correlation coefficient of smoothed density fields at an epoch when $k_{\rm NL} = 32k_f$ is shown plotted versus rms density contrast of the $N$–body simulation (bottom panels). The left-hand (right-hand) panels correspond to the $n = 2$ ($n = 0$) model. Both the $N$–body simulation and the given approximation are smoothed with the same smoothing length to yield a given data point.

Fig. 3.— Evolution of the number of clumps (top panels) and the total mass in clumps (bottom panels) for the $n = 2$ model (left-hand panels) and for the $n = 0$ model (right-hand panels). (We only count those clumps that contain at least 0.1 per cent of the total mass.) In defining clumps we use density thresholds $\rho_c = \rho_0$ and $\rho_c = 2\rho_0$, respectively, for the $n = 2$ and $n = 0$. (Here $\rho_0$ is the average density.)

Fig. 4.— Probability density function is shown plotted for the $n = 2$ model (left-hand panel) and for the $n = 0$ model (right-hand panel) at an epoch when $k_{\rm NL} = 16k_f$. Note that even in the $n = 0$ case, wherein the potential does not evolve appreciably, SP does show a slight improvement over FP.